\documentclass[lettersize,journal]{IEEEtran}
\usepackage{amsmath,amsfonts}
\usepackage{amssymb}
\usepackage{algorithm}
\usepackage{algorithmic}
\usepackage{array}
\usepackage[caption=false,font=normalsize,labelfont=sf,textfont=sf]{subfig}
\usepackage{textcomp}
\usepackage{stfloats}
\usepackage{color}
\usepackage{url}
\usepackage{verbatim}
\usepackage{cite}
\usepackage{mathrsfs}
\usepackage{bbding}
\usepackage{fontawesome}
\usepackage{subcaption}
\usepackage{pifont}
\usepackage{graphicx}
\usepackage{subfig}
\usepackage{booktabs}
\usepackage{multirow}

\begin{document}

\title{Breaking Near-Field Communication Barriers: Focused, Curved, or Airy Beamforming?}

\author{
\IEEEauthorblockN
{
Shupei Zhang, \IEEEmembership{Graduate Student Member, IEEE}, Boya Di, \IEEEmembership{Senior Member, IEEE},\\ and Lingyang Song, \IEEEmembership{Fellow, IEEE}
}
\thanks{Shupei Zhang, Boya Di and Lingyang Song are with State Key Laboratory of Photonics and Communications, School of Electronics, Peking University, Beijing, China (emails: zhangshupei@pku.edu.cn; diboya@pku.edu.cn; lingyang.song@pku.edu.cn).

Lingyang Song is also with School of Electronic and Computer Engineering, Peking University Shenzhen Graduate School, Shenzhen 518055, China.
}
}
\maketitle

\begin{abstract}
To meet the requirements for high data rates and ubiquitous connectivity in 6G networks, higher frequencies and larger array apertures are employed to enhance spatial resolution and spectral efficiency.
This evolution leads to an expansion of the near-field region, where spherical-wave focusing can significantly enhance received power.
However, the pervasive presence of obstacles in near-field environments makes communication in obstructed scenarios a critical challenge, particularly for sensitive high-frequency links with high penetration losses.
In this paper, we propose a new waveform, termed the \emph{near-field Airy beam}, which is tailored to the amplitude and phase characteristics of obstructed near-field channels.
By integrating non-uniform amplitude response with non-linear phase profile, the proposed Airy beam forms specific curved trajectories, energy distributions, and focal points, enabling energy concentration at the user even after circumventing obstacles. An Airy beamforming algorithm is also developed for hybrid beamformer architectures. Considering practical conditions with unknown obstacle and user locations, we design an Airy beam codebook and a low-overhead hierarchical search scheme to identify the optimal user-aligned beam. Simulation results demonstrate that in obstructed environments, the near-field Airy beam achieves a received power gain of over 3 dB compared to conventional waveforms like focused and curved beams, closely approaching the theoretical upper bound. Across the mmWave to THz bands and various obstacle dimensions, the proposed beam training scheme consistently outperforms traditional methods in terms of spectral efficiency while maintaining a comparable training overhead.
\end{abstract}


\section{Introduction}

With the evolution toward 6G wireless networks, the demand for extreme data rates and ubiquitous connectivity has reached unprecedented levels~\cite{6G1,6G2}. To satisfy these stringent performance requirements, 6G is expected to exploit higher frequency bands and employ large-scale arrays to enhance spatial resolution for boosting spectral efficiency~\cite{6G3,6G4}. The simultaneous increase in antenna aperture size and operating frequency leads to a significant expansion of the Rayleigh distance of transceivers, thereby shifting the operating environment of communication scenarios from the conventional far-field to the near-field region~\cite{NF1,TCCN}.
Specifically, for an antenna with a $0.4\text{ m}$ aperture operating at $100\text{ GHz}$, the near-field boundary extends to $106.7\text{ m}$.

In such a near-field region, the traditional far-field assumption of plane-wave propagation is no longer valid. Instead, the electromagnetic~(EM) wavefront could be modeled as a spherical wave, characterized by a non-linear phase distribution across the antenna array~\cite{NF2}. This fundamental shift from linear to non-linear phase control enables a new paradigm known as \emph{beam focusing}, which allows the EM energy to be concentrated into a specific spatial point rather than just a direction~\cite{NF3}. By exploiting this additional distance-dimension degree of freedom, near-field communication can substantially enhance the received signal gain and mitigate spatial interference, offering a potential for improving the spectral efficiency of 6G systems.

Despite the significant gain provided by near-field beam focusing, the reliability of such high-precision links is inherently vulnerable to environmental dynamics, particularly the ubiquitous presence of obstacles. Specifically, the high-frequency signals suffer from severe insertion and penetration losses when encountering common materials, making the links extremely sensitive to even minor obstacles~\cite{block1,block2}. What's more, unlike the broader angular coverage of far-field beams, near-field focused beams are characterized by an extremely fine spatial distribution~\cite{block3}. While this precision enhances energy efficiency, it also renders the focal spot highly susceptible to blockage, where a small obstacle at a critical position can completely intercept the EM wave propagation. Consequently, the performance of conventional near-field focused beamforming schemes degrades markedly in blockage-prone environments, leading to frequent link outages.

To circumvent communication disruptions caused by obstacles, prior researches have explored specialized waveform designs for obstructed environments, primarily focusing on Airy beams~\cite{Airy1,Airy2,Airy3,Airy4} and curved beams~\cite{Curved1,Curved2,Curved3,Curved4,Curved5,Curved6}. Specifically, classic Airy beams are a type of wavepackets whose initial field distribution conforms to an Airy function. These beams are characterized by their unique propagation properties, including self-bending trajectories and self-healing capabilities.
Curved beams are generated by manipulating the phase profile of the antenna array to direct EM waves along specific curved trajectories, thereby bypassing obstacles.
A typical curved beam is designed based on the Fourier spectrum of the Airy beams to leverage its self-acceleration property.
Building on this principle, \cite{Curved4} integrates a cubic curvature phase term into the near-field focused beam profile to realize such curved beams with controllable parabolic trajectories.
To obtain the optimal curved beam aligned with the target user, \cite{Curved1} designs a two-stage curved beam search scheme to reduce the scanning overhead. Meanwhile, \cite{Curved2} proposes a physics-informed neural network based framework for the real-time optimization of curved beams to bypass obstacles in sub-THz wireless networks.
By manipulating the phase profile over the transmitting aperture, \cite{Curved3} regulates the launching direction of individual rays so that their envelope follows a pre-defined curved trajectory.
The authors in~\cite{Airy1} propose a secure transmission scheme where the message is delivered via diverse wavefronts including Airy beams, thereby enhancing data security.
In~\cite{Airy2}, two-dimensional Airy beams are synthesized using metasurfaces, validating their fundamental non-diffracting, self-bending, and self-healing properties.

However, existing waveforms, including focused, curved, and classical Airy beams, may not comprehensively account for the intricate characteristics of obstructed near-field channels.
Specifically, while the non-linear phase of focused beams can concentrate energy at the user in unobstructed environments, their focusing performance significantly deteriorates under blockage.
Curved beams offer benting trajectories to bypass obstacles and mitigate power loss, but the customization of such trajectories often incurs energy attenuation before reaching the user. While classical Airy beams utilize amplitude response modulation based on the Airy function to achieve self-bending and self-healing, they fail to leverage the non-linear phase distribution inherent in near-field channels, which prevents the effective focusing of beam energy at the user.

To better adapt to obstructed near-field channels, this paper proposes a near-field Airy beam by simultaneously incorporating the amplitude profile of classical Airy beams and the phase distribution of focused beams. The core of this waveform design is two-fold: \textbf{1) Amplitude Matching:} The oscillating decay of the Airy function, characterized by a dominant main lobe and suppressed sidelobes, aligns with the non-uniform amplitude response of blocked channels, where unobstructed links are strong and obstructed ones are weak. This ensures an effective amplitude match between the waveform and the channel.  \textbf{2) Phase Focusing:} The integration of a non-linear phase profile leverages the dual angle-distance degrees of freedom of the near field. This allows the beam to focus energy on the user even after circumventing obstacles. After that, a beam training scheme with unknown obstacle and user locations is designed. Moreover, the Airy beamforming for the widely used hybrid beamformer architecture is presented. Compared to conventional schemes, the proposed method substantially enhances the received power without additional hardware or increased training overhead.

The contributions of this paper are summarized as follows.
\begin{itemize}
\item To cope with near-field communications under blockage, the near-field Airy beam is proposed by integrating the Airy function-based amplitude response with the near-field phase profile.
Specific guidelines for generating such beams are provided, which are governed by four key parameters, and their impacts on the resulting beam pattern are discussed.
Benefiting from its superior matching with obstructed channel responses, the proposed Airy beam achieves a received power gain of over 3 dB compared to focused and curved beams, closely approaching the theoretical upper bound. Moreover, an Airy beamforming for hybrid architectures is developed, which achieves performance close to that of a fully digital structure.

\item Considering practical conditions where both obstacle and user locations are unknown, we design a specialized Airy beam codebook and a corresponding beam training scheme to identify the optimal beam alignment. To maximize the unique coverage benefits of Airy beams, we derive sampling criteria for the beam parameters to construct the Airy beam codebook. Additionally, a hierarchical Airy beam search scheme is devised to maintain high gain while keeping the training overhead at the same level as focused and curved beam methods.

\item Simulation results demonstrate that the proposed near-field Airy beam effectively matches both the amplitude and phase responses of obstructed channels, proving it to be a suitable waveform for approaching the theoretical limit in blocked environments.
Specifically, under blockage conditions, the proposed Airy beam training scheme yields average received power improvements of $3.3\text{ dB}$ and $2.5\text{ dB}$ over focused and curved beams at 100 GHz, respectively. Moreover, the performance of the hierarchical search scheme closely approaches that of exhaustive search, effectively avoiding the escalation of training complexity while sustaining high performance gains.

\end{itemize}

\emph{Organization:} The remainder of this paper is organized as follows. Section~\ref{SecII} establishes the EM propagation model in obstructed environments and formulates the received power maximization problem. Section~\ref{SecIII} characterizes the generation of near-field Airy beams, analyzes the impact of beam parameters on the resulting beam patterns, and evaluates the performance of different waveforms under blockage. Section~\ref{SecIV} is dedicated to the design of the Airy beam codebook and the discussion of parameter sampling criteria. Subsequently, an Airy beamforming algorithm for hybrid architectures is proposed, along with a low-overhead hierarchical beam training scheme. Finally, simulation results and conclusions are presented in Section~\ref{SecV} and Section~\ref{SecVI}, respectively.

\emph{Notations:} In this paper, bold lowercase and uppercase letters denote vectors and matrices, respectively. The operators $(\cdot)^T$ and $(\cdot)^H$ represent the transpose and conjugate transpose , while $\|\cdot\|_2$ denotes the $l_2$-norm. The notations $\mathcal{F}\{\cdot\}$ and $\mathcal{F}^{-1}\{\cdot\}$ refer to the Fourier and inverse Fourier transforms, and $\mathcal{Q}(\cdot, B)$ indicates $B$-bit phase quantization to a discrete set.

\section{System Model}\label{SecII}

In this section, we introduce the transmitter and receiver configurations, the EM wave propagation in obstructed environments, and the problem formulation.

\subsection{Scenario Description}

We consider a large-scale array assisted near-field communication system, where the base station~(BS) equipped with a uniform linear array~(ULA) serves a single-antenna user. The ULA consists of $N$ antenna elements with an inter-element spacing of $d$, resulting in a total aperture size of $(N-1)d$. Without loss of generality, the ULA is centered at the origin and aligned along the $y$-axis. Consequently, the Cartesian coordinate of the $n$-th antenna element is given by $[x_n, y_n] = [0, n d]$, where the antenna index is $n \in \{-\frac{N-1}{2}, \dots, 0, \dots, \frac{N-1}{2}\}$.
The antenna beamforming vector is defined as $\mathbf{w} \in \mathbb{C}^{N \times 1}$, where $w_n$ denotes the response of the $n$-th antenna element.
The location of the user is represented by $ [x_u, y_u]$. Due to the presence of obstacles in the propagation environment, the line-of-sight (LoS) links between specific antenna elements and the user may be blocked.
As the operating frequency scales up to the mmWave or even THz bands, non-negligible penetration loss occurs, and even minor obstacles can disrupt the communication.
Therefore, it is imperative to explore near-field transmission schemes under blockage.

\begin{figure}[t]
\setlength{\abovecaptionskip}{0pt}
\setlength{\belowcaptionskip}{0pt}
	\centering
    \includegraphics[width=0.9\linewidth]{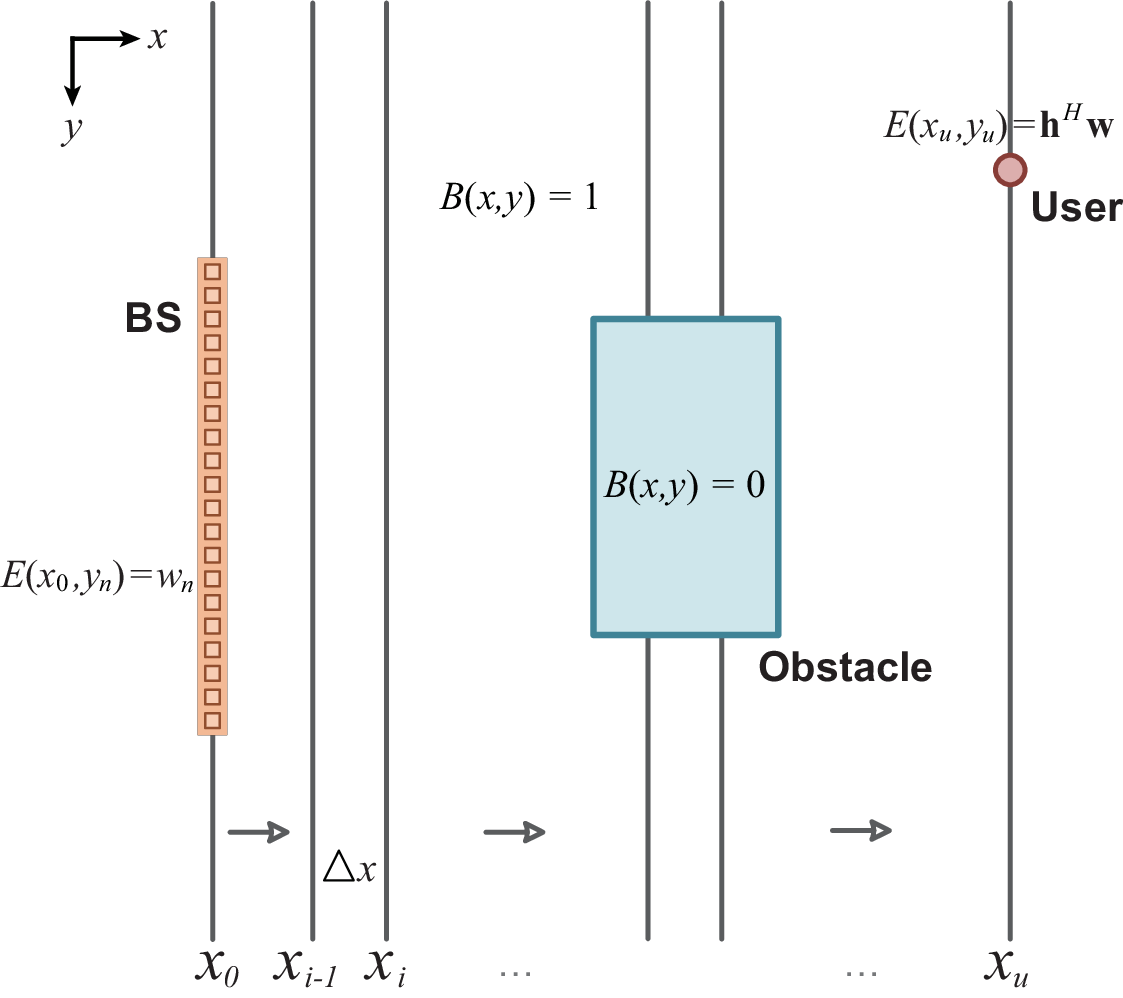}
	\caption{Electric field propagation from BS to user.}
	\label{Fig:system}
\vspace{0em}
\end{figure}

\subsection{Obstructed EM Wave Propagation}

In the presence of obstacles, conventional geometric channel models become inadequate as they fail to characterize the diffraction of EM waves~\cite{Curved1,Curved2}.
In addition, existing channel estimation schemes may not be applicable in such obstacled environments.
According to the Huygens-Fresnel principle, every point on an EM wavefront can be regarded as a secondary source of wavelets. At any subsequent instant, the envelope of these secondary wavelets constitutes the newly formed wavefront~\cite{principle1}. Upon encountering an obstacle, the electric field of the secondary sources within the obstructed region is nullified, thereby disrupting the continuous propagation of the wave.
The propagation process can be accurately characterized by the Rayleigh-Sommerfeld diffraction integral.

As shown in Fig~\ref{Fig:system}, we assume a rectangular obstacle that covers a spatial region $\mathcal{C} = [x_{left}, x_{right}] \times [y_{down}, y_{up}]$.
Given the initial electric field distribution of the antenna, to evaluate the electric field $E(x_u, y_u)$ at the user location, we employ an iterative propagation approach with a spatial step $\Delta x$. Specifically, we start from the initial field at the antenna plane ($x_0=0$), i.e., $E(x_0, y_n)=w_n$. The electric field is computed sequentially at discrete planes $x_i = (i-1) \Delta x$ until the user distance $x_u$ is reached, for $i=2, 3, \dots, I$. To account for the impact of obstructions, a binary matrix $B$ is introduced to indicate whether a specific spatial point is located within the obstacle. In the $i$-th iteration, the electric field at the plane $(x_i, y)$ is obtained by the superposition of all secondary sources from the preceding plane $x_{i-1}$, which can be expressed as~\cite{Curved4,channel1}
\begin{equation}
E(x_i, y) = B(x_i, y) \int E(x_{i-1}, y') \frac{e^{-j \kappa r_u}}{2 \pi r_u^2} x_i \left( j \kappa + \frac{1}{r_u} \right) dy',
\end{equation}
where $\kappa = 2\pi/\lambda$ is the wavenumber, $r_u = \sqrt{(y-y')^2 + \Delta x^2}$ denotes the distance between the source point $(x_{i-1}, y')$ and the observation point $(x_i, y)$.
If the point $(x_i, y)$ is located within the obstacle, i.e., $(x_i, y) \in \mathcal{C}$, the binary index $B(x_i, y) = 0$, otherwise $B(x_i, y) = 1$.

To reduce the computational time, the angular spectrum method (ASM) can be employed to transform the integral calculations into the spatial frequency domain using Fourier transforms. By interpreting the Rayleigh-Sommerfeld integral as a convolution, its representation in the Fourier domain is given by~\cite{channel1}
\begin{equation}
\mathcal{F}\{E(x_i, y)\} = \mathcal{F}\{E(x_{i-1}, y)\} \times H(\kappa_y).
\end{equation}
The notation $\mathcal{F}\{\cdot\}$ denotes the Fourier transform, and $H(\kappa_y)$ represents the transfer function expressed as
\begin{equation}
H(\kappa_y) = \exp \left( -j \Delta x \sqrt{\kappa^2 - \kappa_y^2} \right),
\end{equation}
where $\kappa_y$ is the spatial frequency along the $y$-axis. Finally, by applying the inverse Fourier transform, the electric field at $(x_i, y)$ can be efficiently computed as$$E(x_i, y) = B(x_i, y) \cdot \mathcal{F}^{-1} \left\{ \mathcal{F}\{E(x_{i-1}, y)\} \cdot H(\kappa_y) \right\},$$where $\mathcal{F}^{-1}\{\cdot\}$ denotes the inverse Fourier transform.

Based on the ASM iterative calculation of electric fields, we can establish the relationship between this EM approach and the equivalent channel in conventional wireless communications. For a given beamforming vector $\mathbf{w}$, the electric field $E(x_u, y_u)$ at the user is fundamentally the superposition of the initial aperture field after propagating through a complex EM environment, expressed as$$E(x_u, y_u) = \mathbf{h}^H \mathbf{w},$$where $\mathbf{h} \in \mathbb{C}^{N \times 1}$ represents the equivalent channel vector between the user and BS in an obstructed scenario. Specifically, the $n$-th entry of $\mathbf{h}$ corresponds to the complex electric field observed at the user coordinates when only the $n$-th antenna element is excited. In contrast to traditional geometric channel models, the channel vector $\mathbf{h}$ obtained by the ASM is not a mere superposition of phase delays. Instead, it captures the diffraction effects arising from the truncation of secondary wave sources upon encountering obstacles~\cite{channel1}.

\subsection{Problem Formulation}

Our objective is to maximize the received power at the user by configuring the antenna beamformer $\mathbf{w}$.
This is equivalent to maximizing the electric field intensity $|E(x_u, y_u)|$ at the user's location by adjusting the initial electric field profile $E(x_0, y_n)$ on the antenna aperture.
However, since the positions of both the user and the obstacles are unknown, and the resulting EM propagation environment is highly intricate, the optimal initial electric field cannot be determined in advance. In such a scenario, beam training is a widely adopted beamforming scheme~\cite{codebook1}. Specifically, a codebook $\mathcal{W} = \{\mathbf{w}_1, \mathbf{w}_2, \dots, \mathbf{w}_K\}$ is pre-designed, where each codeword $\mathbf{w}_k$ represents a specific antenna beamformer that corresponds to a unique initial electric field, i.e., $E(x_0, y_n)=[\mathbf{w}_k]_n$~\cite{Curved1}. During the training process, the BS sequentially transmits signals using each codeword, and the user finally feeds back the index of the codeword that yields the maximum electric field intensity. This beam training problem can be formulated as~\cite{Curved2}
\begin{equation}
\begin{gathered}
    k^* = \arg \max_{\mathbf{w}_k \in \mathcal{W}} |\mathbf{h}^H \mathbf{w}_k| \\
    \Updownarrow \\
    k^* = \arg \max_{\mathbf{w}_k \in \mathcal{W}} |E(x_u, y_u | \mathbf{w}_k)| \\
    \text{s.t. } \|\mathbf{w}_k\|_2^2 = P, \quad \forall k
\end{gathered}
\end{equation}
where $E(x_u, y_u| \mathbf{w}_k)$ denotes the electric field at the user location $(x_u, y_u)$ generated by the $k$-th codeword.
The constraint $\|\mathbf{w}_k\|_2^2 = P$ ensures that each codeword satisfies the transmit power requirement.

In contrast to conventional near-field codebook design and beam training approaches that rely on LoS propagation, the direct links between individual antenna elements and the user may be occluded by obstacles in the blockage case. Relying on such LoS-oriented codebooks may lead to significant beam blockage and potential link outages. Consequently, it is imperative to develop blockage-resilient codebooks to maintain robust connectivity in obstructed environments.

\section{Generation and Characterization of Airy Beams}\label{SecIII}

In this section, we first review the focused and curved beams, and then propose the near-field Airy beams. Finally, the performance of various waveforms in obstructed environments is demonstrated.

\begin{table*}[t]
\centering
\caption{Comparison of Different Waveforms}
\label{tab:beam}
\begin{tabular}{@{}ccccc@{}}
\toprule
\textbf{Waveform} & \textbf{Configuration} & \textbf{Amplitude} & \textbf{Phase} & \textbf{Implementation} \\ \midrule
Steered Beam & $w_{\text{Steer}, n}(\theta) =\sqrt{\frac{P}{N}} \exp \left( -j \kappa  nd \sin \theta  \right)$ & Uniform & Linear & Analog/Digital/Hybrid  \\ \addlinespace
Focused Beam & $w_{\text{Focus}, n}(\theta, r) =\sqrt{\frac{P}{N}} \exp \left( j \kappa \left( -nd \sin \theta + \frac{n^2 d^2 \cos^2 \theta}{2r} \right) \right)$ & Uniform & Non-linear & Analog/Digital/Hybrid  \\ \addlinespace
Curved Beam & $w_{\text{Curve}, n}(\theta, r, c) = w_{\text{Focus}, n}(\theta, r) \cdot \exp \left( -j  \frac{(2\pi c)^3 (nd)^3}{3} \right)$ & Uniform & Non-linear & Analog/Digital/Hybrid \\ \addlinespace
Classic Airy Beam & $\tilde{w}_{\text{Airy}, n} = {Ai}\left( \frac{nd}{s} \right) \exp\left( a \frac{nd}{s} \right)$ & Non-uniform & $0/ \pi$ & Digital/Hybrid \\ \addlinespace
Near-Field Airy Beam & $w_{\text{Airy}, n}(\theta, r, s, a)=\frac{1}{\mathcal{N}} \tilde{w}_{\text{Airy}, n} \cdot {\exp \left( j \kappa \left( -nd \sin \theta + \frac{n^2 d^2 \cos^2 \theta}{2r} \right) \right)}$ & Non-uniform & Non-linear & Digital/Hybrid \\ \bottomrule
\end{tabular}
\vspace{5pt}
\end{table*}

\subsection{Review of Focused Beams and Curved Beams}
\textbf{Focused Beams:}
With the escalation of carrier frequencies and the expansion of antenna array scales, the near-field region of the BS markedly enlarges. In this region, the channel is characterized by both the distance and angle of the user, necessitating a paradigm shift in beamforming from far-field plane-wave models to near-field spherical-wave models~\cite{focus1}. Specifically, near-field spherical waves, also referred to as focused beams, enable EM energy to be concentrated at a specific point rather than being dispersed in a single direction, thereby substantially enhancing the received signal gain~\cite{focus2}. Assuming the user is located at polar coordinates $(\theta, r)$, the response of the $n$-th antenna element for a focused beam is given by
\begin{equation}\label{focus}
w_{\text{Focus}, n}(\theta, r) = \sqrt{\frac{P}{{N}}} \exp \left( j \kappa \left( -y_n \sin \theta + \frac{y_n^2 \cos^2 \theta}{2r} \right) \right),
\end{equation}
where the phase term is obtained by the second-order Taylor expansion of the exact distance between the $n$-th antenna element and the focal point, i.e., $r_n = \sqrt{x_u^2 + (y_u - y_n)^2}$.

\textbf{Curved Beams:}
While focused beams are capable of concentrating EM energy at a specific point in the near-field, their propagation trajectories follow straight lines. Consequently, when the LoS path between the BS and the user is blocked by obstacles, the received energy at the user significantly degrades. To address this limitation, curved beams have been proposed, which exhibit a parabolic trajectory that allows the EM wave to circumvent obstacles and enhance the received gain. Building upon the focused beams, the response of the $n$-th antenna element required to generate such a curved trajectory is given by~\cite{Curved1,Curved2}
\begin{equation}\label{Curve}
w_{\text{Curve}, n}(\theta, r, c) = w_{\text{Focus}, n}(\theta, r) \cdot \exp \left( -j  \frac{(2\pi c y_n)^3}{3} \right),
\end{equation}
where $c$ is the curvature coefficient that governs both the direction and the degree of the trajectory's bend. Specifically, the sign of $c$ determines the direction of the curvature (e.g., upward or downward), while the magnitude of $|c|$ dictates the intensity of the self-acceleration effect.

\begin{figure*}[t]
\setlength{\abovecaptionskip}{0pt}
\setlength{\belowcaptionskip}{0pt}
	\centering
    \includegraphics[width=\linewidth]{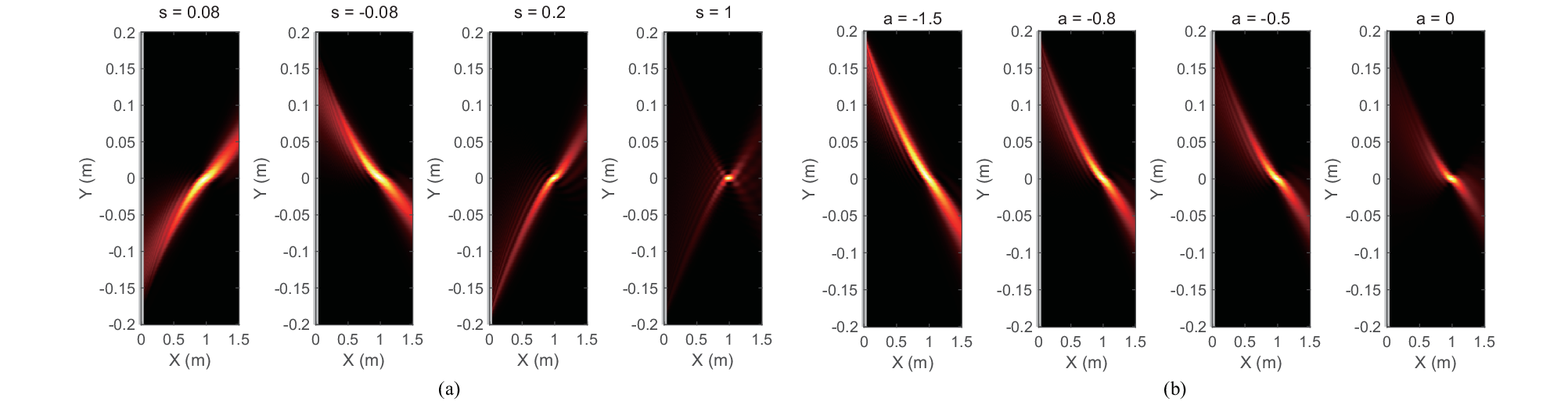}
	\caption{Influence of Airy beam parameters on beam patterns: (a) impact of $s$ for a fixed $a = -0.8$, and (b) impact of $a$ for a fixed $s = -0.1$.}
	\label{Fig:impact}
\vspace{0em}
\end{figure*}

\subsection{Near-Field Airy Beams}

While curved beams provide the possibility to circumvent obstacles, their electric field distribution is determined by the curved trajectory itself rather than specifically tailored for the obstructed near-field channels.
To generate such a curved trajectory, EM waves from certain antenna elements may dissipate before reaching the user, leading to a limited enhancement in received power compared to focused beams. To further improve the received gain after bypassing obstacles, the Airy beam is formed by a non-uniform electric field distribution~\cite{Airy5}. Unlike conventional beams, the Airy beam is a unique wavepacket solution to the paraxial Helmholtz equation. It is characterized by physical properties like non-diffraction and self-acceleration\cite{Airy6}. 
The beam can maintain its intensity profile while its main lobe follows a specific curved trajectory during propagation.

Specifically, the initial field distribution of the Airy beam at the $n$-th antenna element can be expressed as~\cite{Airy1,Airy2}
\begin{equation}
\tilde{w}_{\text{Airy}, n}(s,a) = {Ai}\left( \frac{y_n}{s} \right) \exp\left( a \frac{y_n}{s} \right),
\end{equation}
where $Ai(\cdot)$ denotes the Airy function $Ai(x) = \frac{1}{\pi} \int_{0}^{\infty} \cos \left( \frac{1}{3}t^3 + xt \right) dt$, $s$ represents the spatial scaling factor that governs the curved trajectory, and $a$ is the exponential decay parameter utilized to regulate the energy distribution and satisfy the energy constraint.

To concentrate the energy of the Airy beam at a specific point in the near-field, we integrate the Airy amplitude envelope with the near-field phase term. Consequently, the resulting near-field Airy beam is jointly governed by the spatial scaling factor $s$, the exponential decay parameter $a$, the steering angle $\theta$, and the focusing distance $r$, which is expressed as
\begin{equation}\label{NF_airy}
\small
w_{\text{Airy}, n}(\theta, r, s, a) = \frac{1}{\mathcal{N}} \tilde{w}_{\text{Airy}, n} \cdot {\exp \left( j \kappa \left( -y_n \sin \theta + \frac{y_n^2 \cos^2 \theta}{2r} \right) \right)},
\end{equation}
where $\mathcal{N}$ is the normalization factor ensuring $\|\mathbf{w}\|_2^2 = P$.
Therefore, by setting the amplitude response of $n$-th antenna to $\frac{1}{\mathcal{N}} {Ai}\left( \frac{y_n}{s} \right) \exp\left( a \frac{y_n}{s} \right)$ and its phase response to $\kappa \left( -y_n \sin \theta + \frac{y_n^2 \cos^2 \theta}{2r} \right)$, the specific near-field Airy beam corresponding to~\eqref{NF_airy} can be generated.
For convenience, the term Airy beam hereafter refers to the near-field Airy beam generated by~\eqref{NF_airy}.

Fig.~\ref{Fig:impact} illustrates the impact of $s$ and $a$ on the Airy beam pattern. The system parameters are configured with a carrier frequency $f = 100$ GHz, $N = 266$ elements, spacing $d = \lambda/2$, and total power $P = 5$.
\begin{itemize}
\item \emph{Spatial scaling factor $s$}: As shown in Fig. ~\ref{Fig:impact}(a), for a fixed $a$, the sign of $s$ determines the direction of the beam curvature. The magnitude $|s|$ dictates the degree of bending: a smaller $|s|$ results in more pronounced curvature, while a larger $|s|$ leads to a flatter trajectory until the beam eventually degenerates into a focused beam.
\item \emph{Exponential decay parameter a}: As depicted in Fig. ~\ref{Fig:impact}(b), with $s$ held constant, $a$ regulates the energy distribution of the Airy beam. A larger $|a|$ results in a wider main lobe, whereas a smaller $|a|$ yields a more concentrated energy distribution within the main lobe.
\end{itemize}

\begin{figure*}[h]
\setlength{\abovecaptionskip}{0pt}
\setlength{\belowcaptionskip}{0pt}
	\centering
    \includegraphics[width=0.9\linewidth]{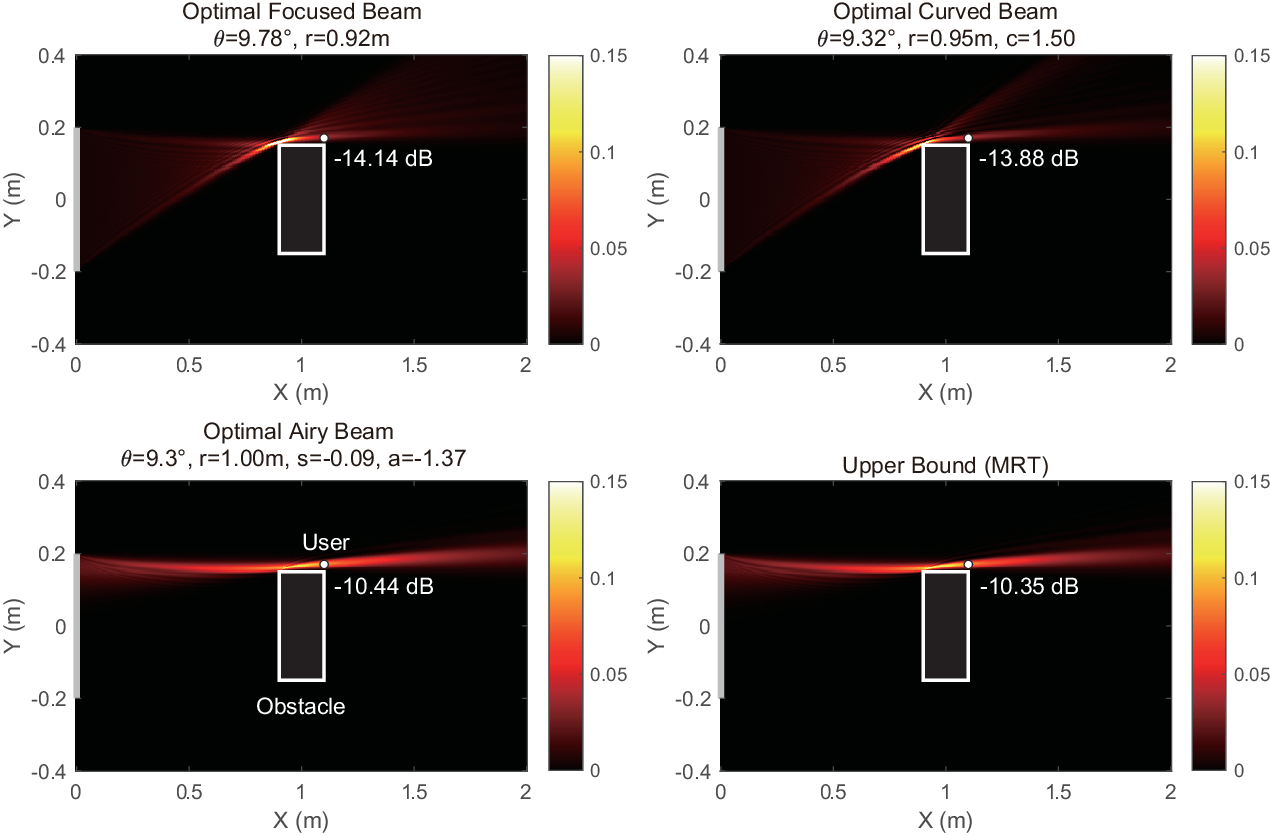}
	\caption{Beam patterns of the optimal focused, curved, Airy, and MRT beams under blockage.}
	\label{Fig:beam pattern}
\vspace{0em}
\end{figure*}

\subsection{Performance of Airy Beams Under Blockage}

In near-field communications under blockage, it is imperative to account for spatial variations in both channel phase and amplitude. In unobstructed environments, the link gains between the user and individual antenna elements are quasi-uniform, enabling focused beams with uniform amplitude profiles to perform optimally. However, in the presence of blockages, the amplitude response becomes highly non-uniform. Blocked links experience severe attenuation, while unblocked links maintain high signal intensity. This spatial discrepancy renders phase-only near-field beamforming strategies inadequate. To mitigate this, the proposed near-field Airy beam not only employs a non-linear phase profile to manage channel phase variations but also utilizes a non-uniform Airy amplitude distribution to adapt to the blockage-induced fluctuations in channel magnitude.
A comparison summary of different beams is presented in Table~\ref{tab:beam}.

Fig.~\ref{Fig:beam pattern} illustrates the obstructed scenario with a carrier frequency of $f = 100$ GHz, $N = 266$ antenna elements, and a transmit power of $P = 5$. The obstacle is within the spatial region $[0.9, 1.1] \times [-0.15, 0.15]$ m, while the user is located at $[1.1, 0.17]$ m. Through fine-grained beam scanning around the user's vicinity, we obtain the optimal beam patterns for the focused, curved, and Airy beams, alongside the theoretical benchmark provided by maximum ratio transmission~(MRT).
The MRT beamformer is designed under the assumption that the channel $\mathbf{h}$ between the antenna array and the user is perfectly known, where the beamformer $\mathbf{w}$ is given by $\mathbf{w} = \frac{\mathbf{h}}{\|\mathbf{h}\|_2} \sqrt{P}$.
As depicted, the focused beam fails to effectively concentrate energy at the user location due to severe blockage. While the curved beam leverages its parabolic trajectory to partially circumvent the obstacle, its main-lobe energy considerably dissipates before reaching the user. In contrast, the proposed near-field Airy beam, benefiting from the joint control of amplitude and phase, enables the signal to propagate along high-gain links that bypass the blockage, thereby successfully concentrating energy at the user with a substantial gain enhancement. Notably, both the beam profile and the received gain of the Airy beam closely approximate those of the theoretically optimal MRT scheme.

\section{Airy Beam Codebook Design and Low-Overhead Search Scheme}\label{SecIV}

In this section, we first propose an exhaustive Airy beam codebook and discuss the sampling criteria for its parameters. Moreover, an Airy beamforming algorithm tailored for hybrid beamformer architectures is developed. Finally, a hierarchical Airy beam search scheme is devised to reduce the training overhead.

\subsection{Airy Beam Codebook}

In practical conditions, the exact locations of both obstacles and users are typically unknown, making it challenging to directly determine the optimal Airy beam to serve the user. While channel estimation in obstructed environments is theoretically viable, it entails prohibitive computational complexity in near-field regions. To address this, it is necessary to design a dedicated Airy beam codebook, which enables the BS to align the beam with the user through beam training.

The design of the Airy codebook is based on the sampling of four key parameters: the steering angle $\theta$, the focusing distance $r$, the spatial scaling factor $s$, and the exponential decay parameter $a$. The codebook is defined as the following set $$\mathcal{W}_{\text{Airy}} = \left\{ \mathbf{w}(\theta, r, s, a) \mid \theta \in \Theta, r \in \mathcal{R}, s \in \mathcal{S}, a \in \mathcal{A} \right\},$$where $\Theta = \{ \theta_1, \theta_2, \dots, \theta_{N_\theta} \}$, $\mathcal{R} = \{ r_1, r_2, \dots, r_{N_r} \}$, $\mathcal{S} = \{ s_{1}, s_{2}, \dots, s_{N_s} \}$, and $\mathcal{A} = \{ a_1, a_2, \dots, a_{N_a} \}$ represent the discrete sampling sets for the steering angle, focusing distance, spatial scaling factor, and exponential decay parameter, respectively.

\subsection{Sampling Criteria for Airy Beam Coefficients}

In conventional far-field DFT codebooks, the design of codewords is typically based on the uniform sampling of the angular domain. In near-field communications, the user channel is jointly determined by both the steering angle and the focusing distance, which motivates the adoption of the polar-domain codebook. To maintain a low correlation between adjacent codewords, the angular domain is sampled uniformly, while the distance domain is sampled non-uniformly. These sampling sets are expressed as\cite{polor}
\begin{equation}
{\Theta} = \left\{ \theta \mid \sin \theta_n = \frac{2n - N - 1}{N}, n = 1, 2, \dots, N \right\},
\end{equation}
\begin{equation}
\mathcal{R} = \left\{ r \mid r_n^m = \frac{Z \cos^2 \theta_n}{ m}, m = 1, 2, \dots, M \right\},
\end{equation}
where $N$ is the number of antenna elements, $M$ denotes the number of sampled distances, and $Z$ controls the spacing of sampling distances at the $n$-th angle. In this polar-domain representation, the distance sampling $r_n^m$ is inversely proportional to the index $m$, ensuring that the codewords are more densely distributed in the region closer to the antenna array where the spherical-wavefront effect is more pronounced.

Regarding the Airy beam codebook, the angular and distance parameters of the phase term are consistent with near-field channel characteristics, thereby adopting the sampling criteria of the polar-domain codebook. We now focus on the sampling criteria for the spatial scaling factor $s$ and the exponential decay parameter $a$ within the Airy function.

\textbf{Exponential decay parameter $a$}: To derive the sampling criterion for $a$, we consider two arbitrary near-field Airy beams focused on the same spatial point, i.e., $r_m = r_n$ and $\theta_m = \theta_n$. Moreover, their spatial scaling factors are assumed to be identical, denoted as $s_{m} = s_{n} = s_0$. Consequently, the focused phase terms of the two beams are perfectly aligned. The correlation between these two beam codewords, $\mathbf{w}(a_m)$ and $\mathbf{w}(a_n)$, can be expressed as the normalized inner product
\begin{equation}
\label{eq:corr_a}
\begin{aligned}
& C(a_m, a_n) = \frac{\left| \mathbf{w}^H(a_m) \mathbf{w}(a_n) \right|}{\| \mathbf{w}(a_m) \|_2 \| \mathbf{w}(a_n) \|_2} \\
&= \frac{\sum_{i=1}^{N} \left| {Ai}\left( \frac{y_i}{s_0} \right) \right|^2 e^{(a_n+a_m)\frac{y_i}{s_0}}}{\sqrt{\sum_{i=1}^{N} \left| {Ai}\left( \frac{y_i}{s_0} \right) \right|^2 e^{2a_m\frac{y_i}{s_0}}} \sqrt{\sum_{i=1}^{N} \left| {Ai}\left( \frac{y_i}{s_0} \right) \right|^2 e^{2a_n\frac{y_i}{s_0}}}}.
\end{aligned}
\end{equation}
It can be observed from \eqref{eq:corr_a} that, when other parameters are fixed, the correlation between the two beams is primarily determined by the linear combination in the exponential decay parameter, i.e., $a_m + a_n$. To ensure consistent coverage of the parameter space, uniform sampling is adopted for $a$. The sampling set for the exponential decay parameter is defined as
\begin{equation}
\small
\mathcal{A} = \left\{ a_k \mid a_k = a_{\min} + \frac{k-1}{N_a - 1}(a_{\max} - a_{\min}), k = 1, 2, \dots, N_a \right\},
\end{equation}
where $[a_{\min}, a_{\max}]$ represents the range of the decay parameter, and $N_a$ is the number of sampling points.

\begin{figure}[t]
\setlength{\abovecaptionskip}{0pt}
\setlength{\belowcaptionskip}{0pt}
	\centering
    \includegraphics[width=0.8\linewidth]{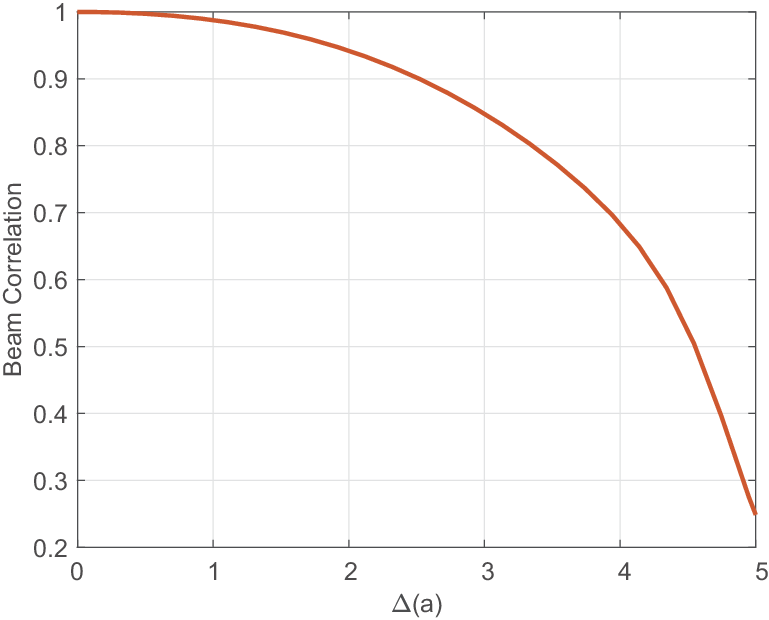}
	\caption{Beam correlation vs. sampling interval of exponential decay parameter  $a$.}
	\label{Fig:a_corr}
\vspace{0em}
\end{figure}

\begin{figure}[t]
\setlength{\abovecaptionskip}{0pt}
\setlength{\belowcaptionskip}{0pt}
	\centering
    \includegraphics[width=0.8\linewidth]{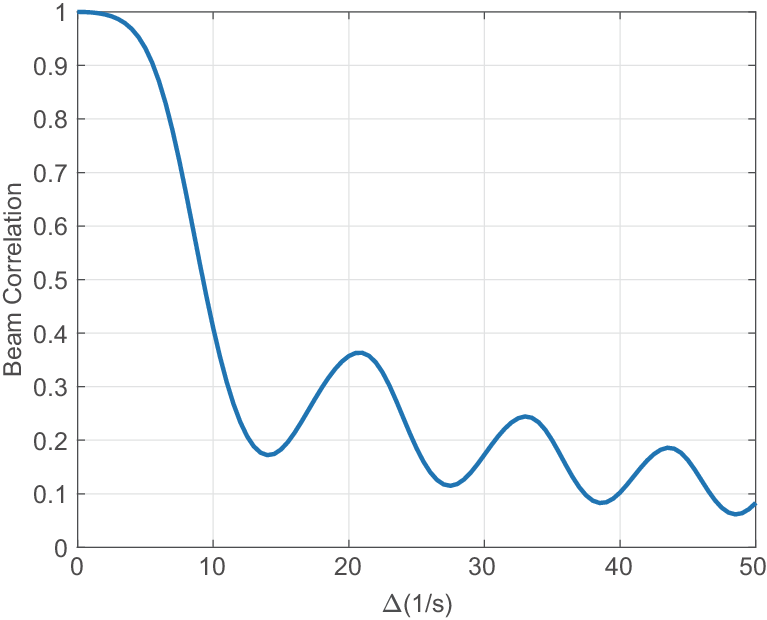}
	\caption{Beam correlation vs. sampling interval of spatial scaling factor $s$.}
	\label{Fig:s_corr}
\vspace{0em}
\end{figure}

\textbf{Spatial scaling factor $s$}: Similarly, to determine the sampling criterion for the spatial scaling factor $w_0$, we consider two arbitrary near-field Airy beams focused on the same location (i.e., $r_m = r_n$ and $\theta_m = \theta_n$) with identical exponential decay parameters ($a_m = a_n = a_0$). Under these conditions, the focusing phase terms cancel out, and the correlation between the two codewords depends solely on the overlap of their amplitude envelopes. The correlation can be expressed as
\begin{equation}\label{eq:corr_s0}
\begin{aligned}
& C(s_m, s_n) = \frac{\left| \mathbf{w}^H(s_m) \mathbf{w}(s_n) \right|}{\| \mathbf{w}(s_m) \|_2 \| \mathbf{w}(s_n) \|_2} \\
&= \frac{\sum_{i=1}^{N} \left| {Ai}\left( \frac{y_i}{s_m} \right) {Ai}\left( \frac{y_i}{s_n} \right) \right| e^{a_0( \frac{y_i}{s_m}+\frac{y_i}{s_n}) }}{\sqrt{\sum_{i=1}^{N} \left| {Ai}\left( \frac{y_i}{s_m} \right) \right|^2 e^{a_0\frac{y_i}{s_m}}} \sqrt{\sum_{i=1}^{N} \left| {Ai}\left( \frac{y_i}{s_n} \right) \right|^2 e^{a_0\frac{y_i}{s_n}}}} \\
&= \sum_{i=1}^{N}  \frac{ \frac{1}{\pi} \left[ \int_{0}^{\infty} \cos\left(\frac{t^3}{3} + \frac{y_i}{s_m}t\right) dt \right] e^{a_0 \frac{y_i}{s_m} }}{\sqrt{\sum_{i=1}^{N} \left| {Ai}\left( \frac{y_i}{s_m} \right) \right|^2 e^{a_0\frac{y_i}{s_m}}}} \times \\
& \hspace{2.5cm}   \frac{ \frac{1}{\pi} \left[ \int_{0}^{\infty} \cos\left(\frac{\tau^3}{3} + \frac{y_i}{s_n}\tau\right) d\tau \right] e^{a_0 \frac{y_i}{s_n} }}{\sqrt{\sum_{i=1}^{N} \left| {Ai}\left( \frac{y_i}{s_n} \right) \right|^2 e^{a_0\frac{y_i}{s_n}}}}
\end{aligned}
\end{equation}
Consider $1/s$ as a spatial frequency scaling factor that dictates the main lobe width and oscillation frequency of the Airy function. Comparing two beams with distinct scaling factors, $s_m$ and $s_n$, essentially evaluates the degree of expansion or compression of the waveforms in the spatial domain. Since the variable involved in the integral kernel is $1/s$ rather than $s$, the spatial frequency components of the waveform parameters should be shifted uniformly to ensure consistent correlation between adjacent codewords.
This relationship implies that to achieve a uniform correlation metric across the codebook, the reciprocal of the spatial scaling factor, i.e., $1/s$, should be uniformly sampled. Accordingly, the sampling set for $s~(s>0)$ is designed as
\begin{equation}
\small
\mathcal{S} = \left\{ s \bigg| \frac{1}{s_{k}} = \frac{1}{s_{\max}} + \frac{k-1}{N_s - 1} \left( \frac{1}{s_{\min}} - \frac{1}{s_{\max}} \right), k = 1, \dots, N_s \right\},\end{equation}
where $[s_{\min}, s_{\max}]$ denotes the range of the spatial scaling factor, and $N_s$ is the number of quantization levels.
To generate Airy beams with diverse curving directions, the aforementioned sampling set is symmetrically extended to the negative domain, i.e., $s < 0$.

\subsection{Airy Beam Codeword Design for Hybrid Beamforming}

The ideal Airy beam denoted as $\mathbf{w} \in \mathbb{C}^{N \times 1}$, assumes a fully digital architecture where both amplitude and phase of each antenna element can be continuously adjusted. However, in practical mmWave and THz systems, the hardware cost and power consumption constrain the use of a fully digital chain~\cite{hybrid}. The hybrid beamforming architecture is typically adopted, connecting $N$ antennas to $N_{\text{RF}}$ RF chains ($N_{\text{RF}} \ll N$).
Our objective is to approximate the ideal Airy beam $\mathbf{w}$ using a hybrid beamformer, which consists of a high-dimensional analog precoder $\mathbf{W}_{\text{RF}} \in \mathbb{C}^{N \times N_{\text{RF}}}$ and a low-dimensional digital precoder $\mathbf{w}_{\text{BB}} \in \mathbb{C}^{N_{\text{RF}} \times 1}$, expressed as $ \mathbf{W}_{\text{RF}} \mathbf{w}_{\text{BB}}$.
This optimization problem is formulated as
$$\begin{aligned}
\min_{\mathbf{W}_{\text{RF}}, \mathbf{w}_{\text{BB}}} & \quad \left\| \mathbf{w} - \mathbf{W}_{\text{RF}} \mathbf{w}_{\text{BB}} \right\|_2^2 \\
\text{s.t.} & \quad |\mathbf{W}_{\text{RF}}(n,m)| = \frac{1}{\sqrt{N}}, \quad \forall n, m, \\
& \quad \angle \mathbf{W}_{\text{RF}}(n,m) \in \mathcal{F}, \\
& \quad \|\mathbf{W}_{\text{RF}} \mathbf{w}_{\text{BB}}\|_2^2 = P ,
\end{aligned}$$
where the first constraint ensures the constant modulus property of the analog phase shifters, the second constraint restricts the phase values to a discrete set $\mathcal{F} = \{ \frac{2\pi k}{2^B} \}_{k=0}^{2^B-1}$, with $B$ denoting the number of quantization bits, and the third constraint limits the power of the hybrid beamformer.

Since the ideal Airy beam is spatially sparse, we reformulate the problem as a sparse reconstruction problem. We employ the orthogonal matching pursuit~(OMP) algorithm to iteratively select the optimal analog steering vectors from a candidate dictionary.
We define the dictionary matrix $\mathbf{A} \in \mathbb{C}^{N \times L}$, which contains $L$ candidate steering vectors. To cover the entire spatial domain, the dictionary is constructed using DFT bases with an oversampling factor $K_{\text{os}}$ as $$\mathbf{A} = [\mathbf{a}(\psi_1), \mathbf{a}(\psi_2), \dots, \mathbf{a}(\psi_L)],$$
$$\mathbf{a}(\psi_l) = \frac{1}{\sqrt{N}} [1, e^{j \pi \psi_l}, \dots, e^{j \pi (N-1) \psi_l}]^T,$$where $\psi_l$ are uniformly sampled from $[-1, 1]$ with $L = K_{\text{os}} N$.
The design of the Airy beam for hybrid beamforming is accomplished in three main steps:
\begin{itemize}
\item \textbf{Step 1:}  In each iteration, the algorithm calculates the correlation between the current residual vector and the dictionary, identifying the column index that yields the maximum projection. The corresponding steering vector is then selected and appended to the analog precoder matrix $\mathbf{W}_{\text{RF}}$. Subsequently, the digital precoder $\mathbf{w}_{\text{BB}}$ is updated via the least squares estimator, followed by an update of the residual error.
\item  \textbf{Step 2:} After the iterative selection of $N_{\text{RF}}$ vectors is completed, the continuous phases of the obtained analog precoder are quantized to the nearest discrete values supported by the $B$-bit resolution of the phase shifters.
\item \textbf{Step 3:} Finally, the digital precoder is recalculated based on the quantized analog beamformer. 
\end{itemize}
The detailed procedure of the proposed Airy beamforming for hybrid beamformer is summarized in Algorithm~\ref{alg:OMP_Airy}.

\begin{algorithm}[t]
\caption{OMP-based Airy Beam Codeword Design for Hybrid Beamforming}
\label{alg:OMP_Airy}
\begin{algorithmic}[1]
\renewcommand{\algorithmicrequire}{\textbf{Input:}}
\renewcommand{\algorithmicensure}{\textbf{Output:}}
\REQUIRE Ideal Airy beam $\mathbf{w}$, number of RF chains $N_{\text{RF}}$, dictionary matrix $\mathbf{A}$, quantization bits $B$.
\STATE \textbf{Initialization:} Residual $\mathbf{r}_0 = \mathbf{w}$, index set $\mathcal{I} = \emptyset$, $\mathbf{W}_{\text{RF}} = []$, iteration counter $t=1$.
\item[] \textbf{OMP-based Beamformer Reconstruction:} 
\WHILE{$t \le N_{\text{RF}}$}
    \STATE Calculate correlation: $\mathbf{p} = \mathbf{A}^H \mathbf{r}_{t-1}$.
    \STATE Select best atom index: $k^* = \arg \max_{k} |\mathbf{p}(k)|$.
    \STATE Update index set: $\mathcal{I} \leftarrow \mathcal{I} \cup \{k^*\}$.
    \STATE Update analog precoder: $\mathbf{W}_{\text{RF}} = [\mathbf{W}_{\text{RF}}, \mathbf{A}(:, k^*)]$.
    \STATE Update digital precoder: 
    $\mathbf{w}_{\text{BB}} = (\mathbf{W}_{\text{RF}}^H \mathbf{W}_{\text{RF}})^{-1} \mathbf{W}_{\text{RF}}^H \mathbf{w}$.
    \STATE Update residual: $\mathbf{r}_t = \mathbf{w} - \mathbf{W}_{\text{RF}} \mathbf{w}_{\text{BB}}$.
    \STATE $t \leftarrow t + 1$.
\ENDWHILE
\item[] \textbf{Phase Quantization:}
\STATE Extract phases: $\mathbf{\Phi} = \angle \mathbf{W}_{\text{RF}}$.
\STATE Quantize phases: $\tilde{\mathbf{\Phi}} = \mathcal{Q}(\mathbf{\Phi}, B)$, where $\mathcal{Q}(\cdot)$ maps phases to the discrete set $\{ \frac{2\pi k}{2^B} \}$.
\STATE Reconstruct quantized analog precoder: $\tilde{\mathbf{W}}_{\text{RF}} = \frac{1}{\sqrt{N}} e^{j \tilde{\mathbf{\Phi}}}$.
\STATE Recalculate digital precoder based on quantized $\tilde{\mathbf{W}}_{\text{RF}}$: \\
$\mathbf{w}_{\text{BB}}^{\star} = (\tilde{\mathbf{W}}_{\text{RF}}^H \tilde{\mathbf{W}}_{\text{RF}})^{-1} \tilde{\mathbf{W}}_{\text{RF}}^H \mathbf{w}$.
\ENSURE Analog precoder $\mathbf{W}_{\text{RF}}$, Digital precoder $\mathbf{w}_{\text{BB}}$.
\end{algorithmic}
\end{algorithm}

\begin{figure}[t]
\setlength{\abovecaptionskip}{0pt}
\setlength{\belowcaptionskip}{0pt}
	\centering
    \includegraphics[width=1\linewidth]{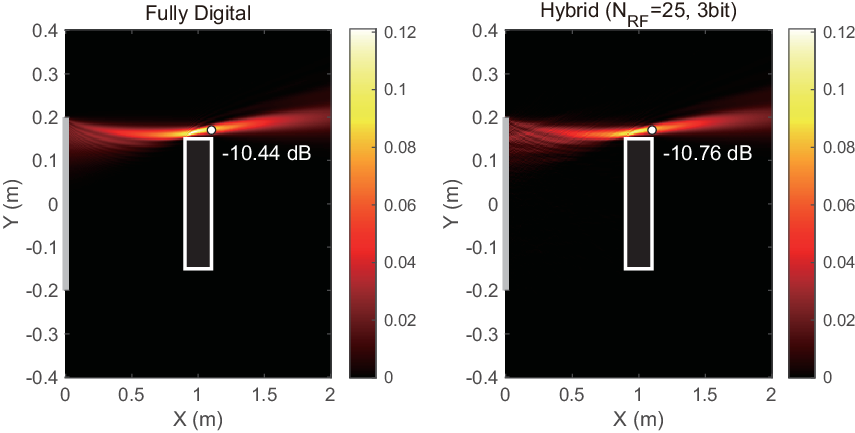}
	\caption{Beam patterns of the fully digital and the proposed hybrid Airy beam codewords.}
	\label{Fig:OMP}
\vspace{0em}
\end{figure}

\begin{figure}[t]
\setlength{\abovecaptionskip}{0pt}
\setlength{\belowcaptionskip}{0pt}
	\centering
    \includegraphics[width=1\linewidth]{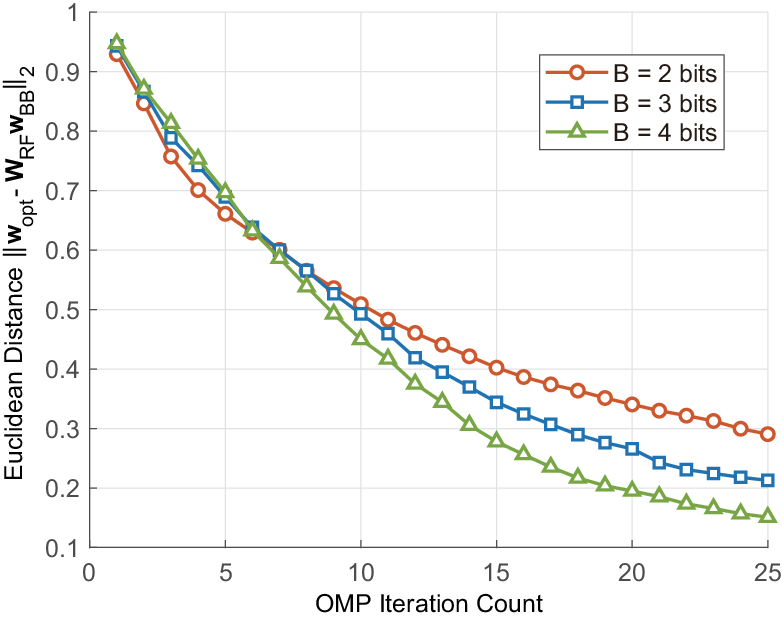}
	\caption{Performance gap between hybrid and fully digital beamformer for Airy beam implementation versus the number of OMP iterations.}
	\label{Fig:ite}
\vspace{0em}
\end{figure}

Fig.~\ref{Fig:OMP} illustrates the beam patterns of the ideal Airy beam generated by the fully digital beamformer and the practical Airy beam generated by the hybrid beamformer. The parameters are configured with a carrier frequency $f = 100$ GHz, $N = 266$ antenna elements, an oversampling factor $K_{\text{os}} = 4$, $N_{\text{RF}} = 25$ RF chains, and $B = 3$ phase quantization bits. It can be observed that the proposed hybrid Airy beam successfully concentrates energy at the user in the presence of the blockage. The beam gain achieved by the hybrid architecture is only $0.32$ dB lower than that of the ideal Airy beam, demonstrating the effectiveness of the proposed scheme in approximating the ideal Airy beam under practical hardware constraints.
Fig.~\ref{Fig:ite} illustrates the performance gap between Airy beams implemented via hybrid and fully-digital beamformers. As the number of OMP iterations increases, the beamforming vector of the hybrid beamformer gradually converges to the ideal Airy beam. The discrepancy relative to the optimal Airy beam $\mathbf{w}_{opt}$ diminishes as the phase quantization resolution $B$ increases. The proposed algorithm reduces the required number of RF chains for the Airy beam from $N$ to $N_{RF}$, and the computational complexity of the OMP-based approach is only $\mathcal{O}(N_{RF})$.

\subsection{Hierarchical Airy Beam Training Scheme}\label{SecIII-D}

While the near-field Airy codebook provides a promising solution for obstructed scenarios by leveraging its unique beam energy distribution and curved trajectories, the introduction of additional Airy parameters leads to an increased search overhead compared to the conventional near-field polar codebook. Specifically, the exhaustive search overhead for the Airy codebook scales as $N_{\theta} \times N_r \times N_s \times N_a$, whereas that of the polar codebook is $N_{\theta} \times N_r$. Therefore, it is imperative to design a low-overhead Airy beam search scheme for practical implementation.

It is noted that the near-field Airy beam is characterized by an Airy function amplitude profile and a focused phase term. Distinct from curved beams, where the focus and curvature components are coupled within the phase distribution, the parameters of the Airy amplitude envelope can be decoupled from the focused phase parameters. This separability enables an independent search of the Airy function parameters and the focused phase, thereby substantially mitigating the training overhead. Accordingly, a hierarchical search scheme is proposed, comprising three sequential stages: phase-focusing, curvature-locking, and energy-distribution as below.
\begin{itemize}
\item \emph{Stage 1: Phase-Focusing}. The near-field polar codebook is employed to search the user’s angle and distance to identify the optimal coordinate pair $(\theta^*, r^*)$.
\item \emph{Stage 2: Curvature-Locking}. Given the fixed optimal coordinates $(\theta^*, r^*)$ and an initial exponential decay parameter $a_0$, the spatial scaling factor $s$ is searched to determine its optimal value $s^*$ in the Airy beam codebook.
\item \emph{Stage 3: Energy-Distribution}. Fixing the optimal coordinates $(\theta^*, r^*)$ and the spatial scaling factor $s^*$, the exponential decay parameter $a$ is scanned to obtain the optimal value $a^*$ in the Airy beam codebook.
\end{itemize}
Upon the completion of the three stages, the optimal Airy beam $\mathbf{w}(\theta^*, r^*, s^*, a^*)$ is obtained. This hierarchical search approach effectively reduces the total training overhead of the Airy codebook from a multiplicative complexity of $N_{\theta} N_r N_s N_a$ to an additive complexity of $N_{\theta} N_r + N_s + N_a$. Consequently, the proposed scheme achieves a search complexity that closely approximates the exhaustive overhead of the polar-domain codebook.

\begin{figure*}[t]
\setlength{\abovecaptionskip}{0pt}
\setlength{\belowcaptionskip}{0pt}
	\centering
    \includegraphics[width=\linewidth]{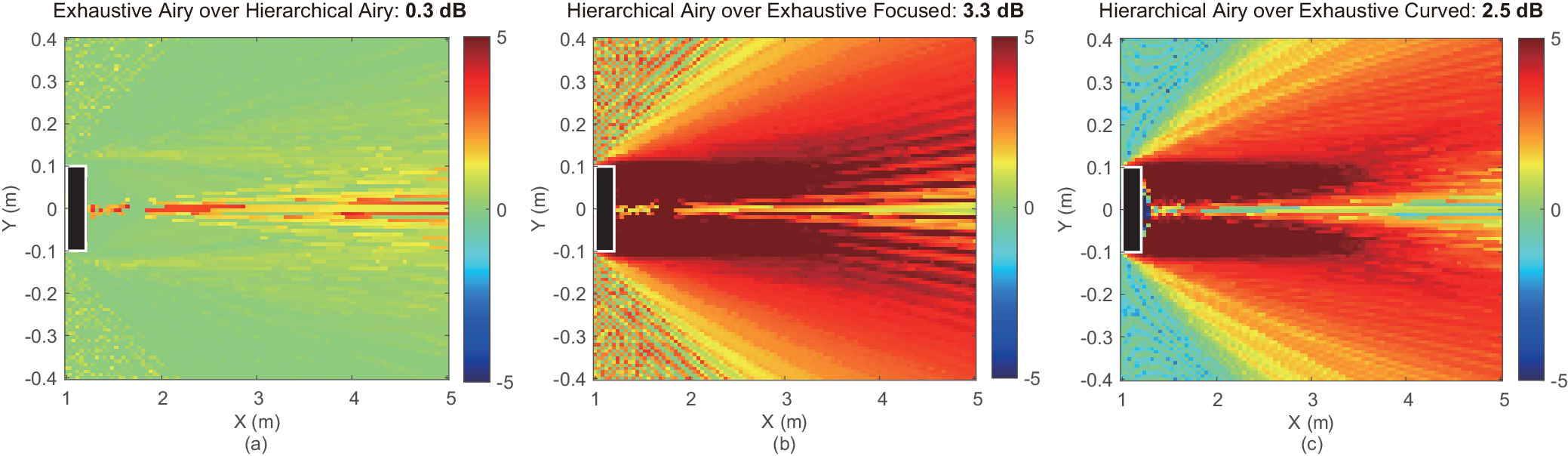}
	\caption{Received power improvement (dB) of (a) exhaustive Airy beam search over hierarchical Airy beam search, (b) hierarchical Airy beam search over exhaustive focused beam search, and (c) hierarchical Airy beam search over exhaustive curved beam search.}
	\label{Fig:improvement}
\vspace{0em}
\end{figure*}

\begin{figure}[t]
\setlength{\abovecaptionskip}{0pt}
\setlength{\belowcaptionskip}{0pt}
	\centering
    \includegraphics[width=0.9\linewidth]{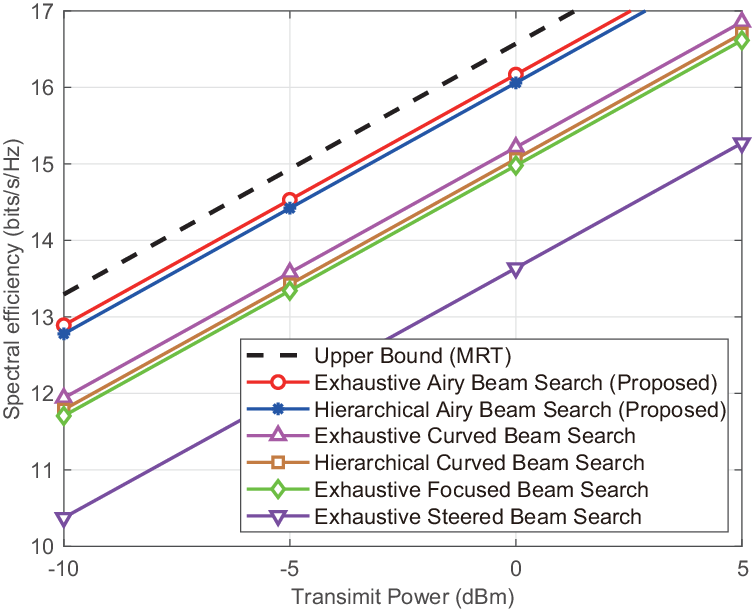}
	\caption{Spectral efficiency vs. transmit power.}
	\label{Fig:SP}
\vspace{0em}
\end{figure}

\section{Simulation Results}\label{SecV}

In this section, we evaluate and compare the performance of the proposed near-field Airy beam and its beam training scheme with conventional benchmarks in obstructed environments. Moreover, the effects of various factors, such as the degree of obstruction, obstacle dimensions, and codebook design, are discussed.

The carrier frequency is set to 100 GHz, and the antenna aperture is 0.4 m, comprising $N = 266$ elements with half-wavelength spacing. For the Airy beam codebook, the angular parameter $\sin\theta$ is uniformly sampled with 90 points in the interval $[\sin(-\pi/4), \sin(\pi/4)]$, while the distance $r$ is non-uniformly sampled with 20 points in $[1, 6]$ m. The exponential decay parameter $a$ is uniformly sampled at 10 points within $[-2, 0]$, and the spatial scaling factor $s$ is non-uniformly sampled at 20 points within $[-0.3, -0.05] \cup [0.05, 0.3]$. The user is randomly distributed in the region $x_u \in [1, 5]$ m and $y_u \in [-0.4, 0.4]$ m, with an obstacle located at $[1, 1.2] \times [-0.1, 0.1]$ m. In the hybrid beamforming architecture, the number of RF chains is 25, and the phase shifter quantization resolution is $B = 3$ bits.

For performance evaluation, the following schemes are considered.
\begin{itemize}
\item \textbf{Perfect CSI (MRT):} 
Assuming the obstructed channel from the antenna array to the user is perfectly known, the maximum ratio transmission beamformer is employed to serve as the performance upper bound.
\item \textbf{Exhaustive Airy Beam Search (Proposed):} 
An exhaustive search is performed over the four-dimensional parameter space of Airy beams. The user feeds back the beam index that maximizes the received signal strength.
\item \textbf{Hierarchical Airy Beam Search (Proposed):} 
This scheme follows the three-stage search procedure proposed in Section~\ref{SecIII-D}, which hierarchically identifies the optimal angle-distance pair $(\theta,r)$, the spatial scaling factor $s$, and the exponential decay parameter $a$.
\item \textbf{Exhaustive Curved Beam Search}~\cite{Curved1,Curved2}: 
An exhaustive search is conducted over the three parameters of curved beams, where the angle and distance sampling are consistent with the Airy codebook, and the curvature $c$ is uniformly sampled at $N_c=21$ points within the interval $[-5, 5]$.
\item \textbf{Hierarchical Curved Beam Search}~\cite{Curved1}:
The angle-distance pair is searched first, followed by a refinement stage to determine the optimal curvature $c$.
\item \textbf{Exhaustive Focused Beam Search}~\cite{polor}:
This scheme traverses a conventional polar-domain codebook to identify the optimal near-field focused beam.
\item \textbf{Exhaustive Steered Beam Search} ~\cite{DFT}:
The DFT codebook is utilized to search for the optimal far-field steered beam.
\end{itemize}

%
%
%
%
%
%

\subsection{Performance Comparison of Beam Training Schemes}

Fig.~\ref{Fig:SP} illustrates the spectral efficiency of various beam training schemes versus the transmit power. Benefiting from the joint regulation of the curved trajectory and energy distribution, the near-field Airy beam achieves a spectral efficiency that most closely approaches the theoretical upper bound. Compared to the linear phase manipulation of steered beams, the non-linear phase control employed by curved and focused beams significantly enhances the spectral efficiency in near-field regions. However, a discernible gap from the theoretical limit remains. The proposed near-field Airy beam further pushes the performance boundaries of near-field beamforming in obstructed environments. Moreover, the proposed hierarchical Airy beam search scheme yields performance nearly identical to that of the exhaustive search while prominently reducing the training overhead. Compared to conventional near-field beam training methods, the hierarchical approach achieves substantial performance gains without a significant increase in training complexity.

Fig.~\ref{Fig:improvement} illustrates the spatial distribution of the received power enhancement achieved by the proposed beam training schemes in obstructed environments. As shown in Fig.~\ref{Fig:improvement}~(a), the exhaustive Airy beam search yields only a 0.3 dB gain improvement in received power over its hierarchical scheme, with the difference primarily concentrated in the signal blind zones behind the obstacle. This confirms that the hierarchical Airy beam search effectively reduces training overhead while preserving transmission performance. Fig.~\ref{Fig:improvement}~(b) and Fig.~\ref{Fig:improvement}~(c) depict the received power improvements of the hierarchical Airy beam search over the exhaustive focused and curved beam search schemes, respectively. It is observed that the hierarchical Airy beam scheme provides substantial power gains of up to 3.3 dB and 2.5 dB respectively, particularly in the obstructed regions. Moreover, the performance gain is more pronounced in areas with higher blockage ratios, demonstrating the superiority of near-field Airy beams in obstructed scenarios.

\subsection{Impact of Blockage Ratio on Received Power}

\begin{figure}[t]
\setlength{\abovecaptionskip}{0pt}
\setlength{\belowcaptionskip}{0pt}
	\centering
    \includegraphics[width=0.9\linewidth]{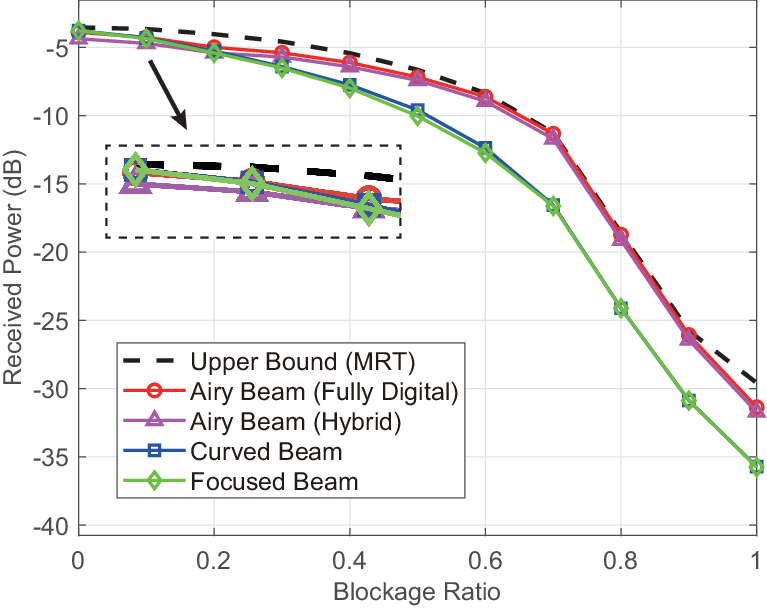}
	\caption{User received power vs. blockage ratio.}
	\label{Fig:block_ratio}
\vspace{0em}
\end{figure}

Fig.~\ref{Fig:block_ratio} illustrates the impact of the user blockage ratio on the received power. The blockage ratio is defined as the proportion of LoS paths between the user and the antenna elements that are blocked by obstacles, where 0 and 1 represent unblocked and fully blocked, respectively. As shown in Fig.~\ref{Fig:block_ratio}, at a blockage ratio of 0, the performances of various beams are similar and all closely approach the upper bound. As the blockage ratio increases, the performance of the Airy beam consistently approximates the upper bound, whereas the performances of the focused and curved beams suffer from degradation.
Benefiting from the self-bending property and the non-uniform energy distribution introduced by the Airy function, the Airy beam is capable of effectively serving users located in obstructed regions. Although the curved beam also possesses bending characteristics, the generation of its curved trajectory sacrifices beam gain, resulting in a performance that is almost identical to that of the focused beam. This demonstrates that the Airy beam is a promising waveform to approach the upper bound in the presence of user blockage.

\begin{figure}[t]
\setlength{\abovecaptionskip}{0pt}
\setlength{\belowcaptionskip}{0pt}
	\centering
    \includegraphics[width=0.9\linewidth]{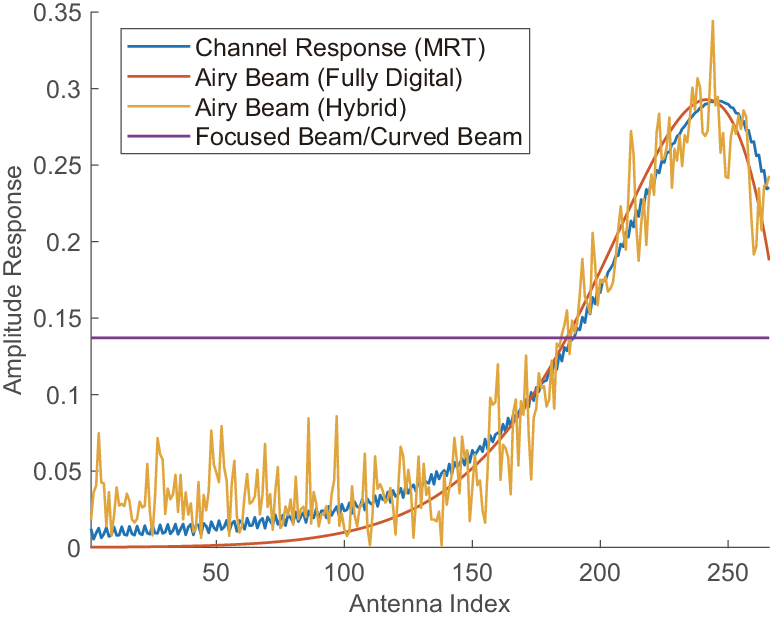}
	\caption{Amplitude response of antenna elements and channels.}
	\label{Fig:amp_response}
\vspace{0em}
\end{figure}

\begin{figure}[t]
\setlength{\abovecaptionskip}{0pt}
\setlength{\belowcaptionskip}{0pt}
	\centering
    \includegraphics[width=0.9\linewidth]{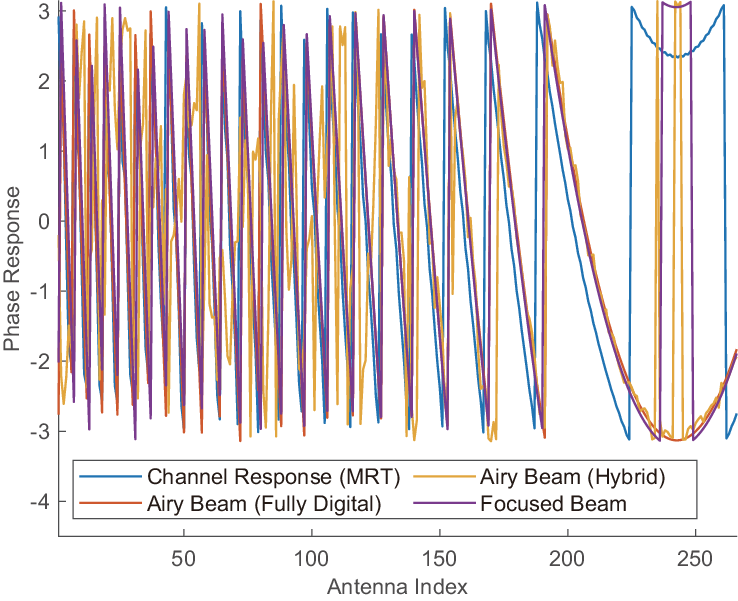}
	\caption{Phase response of antenna elements and channels.}
	\label{Fig:phase_response}
\vspace{0em}
\end{figure}

\begin{figure}[t]
\setlength{\abovecaptionskip}{0pt}
\setlength{\belowcaptionskip}{0pt}
	\centering
    \includegraphics[width=0.9\linewidth]{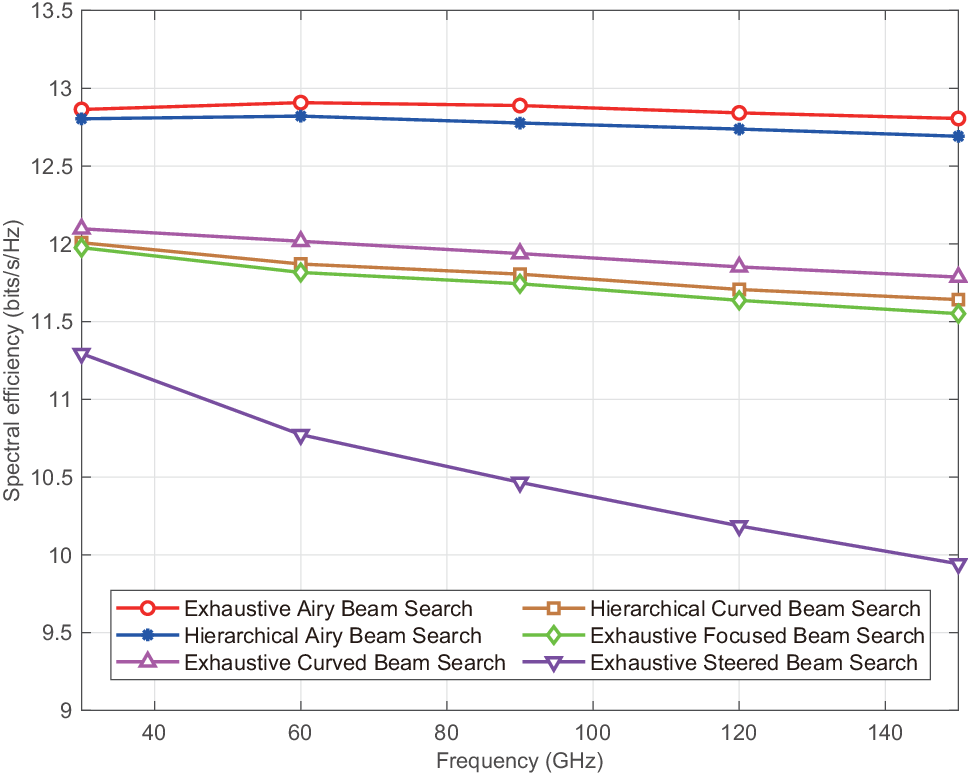}
	\caption{Spectral efficiency vs. carrier frequency.}
	\label{Fig:Fre_SP}
\vspace{0em}
\end{figure}

\subsection{Alignment of Waveforms with Obstructed Channels}

Due to the blockage caused by obstacles, the LoS paths between the user and a subset of antenna elements are interrupted, while the links with other elements remain robust. This disparity results in a non-uniform distribution of channel gain across the antenna array. In this simulation, the obstacle is assumed to be located at $[0.9, 1.1] \times [-0.15, 0.15]$ m, and the user is positioned at $[1.1, 0.17]$ m. The blue line in Fig.~\ref{Fig:amp_response} illustrates the amplitude response of the links between the user and individual antenna elements, clearly exhibiting a non-uniform profile with pronounced variations in link gain. The magnitude response of the Airy beam implemented by the fully digital architecture closely aligns with this non-uniform channel distribution, demonstrating an excellent match. Similarly, the Airy beam via the hybrid architecture also exhibits a comparable non-uniform magnitude profile, albeit with certain fluctuations introduced by the hardware constraints.

In contrast, focused and curved beams, which solely rely on phase control in the codebooks, possess a uniform magnitude distribution that is mismatched for such obstructed scenarios. 
Fig.~\ref{Fig:phase_response} presents the phase distributions of the channel and various waveforms. Given that the user is located in the near-field region, the channel phase exhibits pronounced non-linear characteristics. As shown, all evaluated waveforms employ non-linear phases to achieve a phase match with the channel. In summary, the near-field Airy beam accounts for both the non-linear phase and the non-uniform amplitude of the obstructed user channel, enabling it to outperform conventional waveforms.

\subsection{Beam Training Performance across Frequency}

Given the codebook sampling parameters, antenna aperture, and transmit power, Fig.~\ref{Fig:Fre_SP} illustrates the beam training performance with frequencies. Across the frequency range from typical mmWave (30 GHz) to THz (140 GHz) bands, both exhaustive and hierarchical Airy beam search schemes consistently achieve higher spectral efficiency than the curved and focused beam schemes. This demonstrates the superiority of Airy beams over conventional schemes in blocked scenarios across various frequencies, validating the necessity of simultaneous non-uniform amplitude modulation and non-linear phase control for Airy beams. Moreover, the spectral efficiency of far-field steered beams experiences a noticeable degradation as the frequency increases. This is because near-field effects become increasingly pronounced at higher frequencies, leading to severe phase mismatch for conventional steered beams.

\subsection{The Effect of Codebook Size}

\begin{figure}[t]
\setlength{\abovecaptionskip}{0pt}
\setlength{\belowcaptionskip}{0pt}
	\centering
    \includegraphics[width=0.95\linewidth]{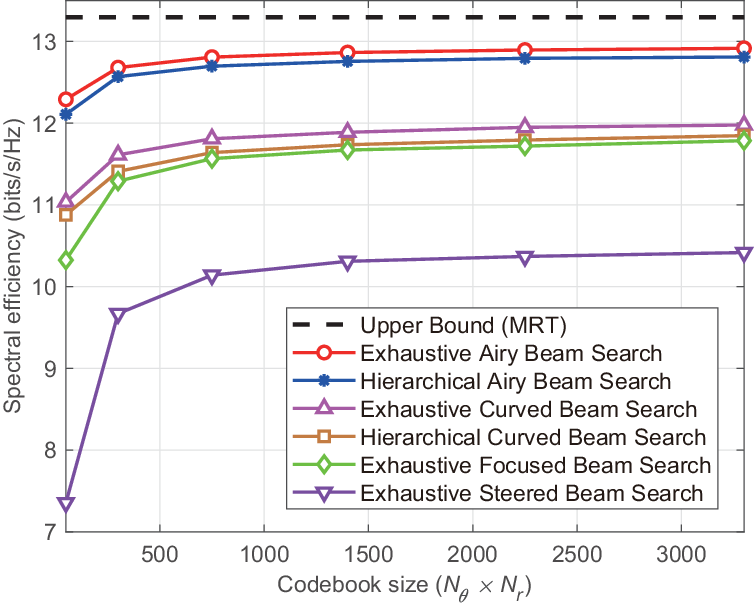}
	\caption{Spectral efficiency vs. codebook size.}
	\label{Fig:booksize}
\vspace{0em}
\end{figure}

The spatial resolution of beam training is dictated by the codebook size, i.e., the density of the codeword sampling. While a higher sampling density of codebook parameters increases the probability of achieving optimal beam alignment with the user, it simultaneously escalates the training overhead~\cite{booksize}. Fig.~\ref{Fig:booksize} illustrates the variation in the user data rate with respect to the number of angular and distance samples. The number of angular samples ranges from 10 to 130 (with an interval of 20), and the number of distance samples varies from 5 to 35 (with an interval of 5). As shown, the data rates for all beam training schemes improve as the sampling density increases.

For all codebook sizes, benefiting from the joint regulation of both the amplitude and phase of EM waves, the data rate of the Airy beam progressively approaches the theoretical upper bound and outperforms both the focused and curved beams. In contrast, the steered beam exhibits the lowest data rate due to significant phase mismatch in the near-field region. As the codebook size increases, the data rates for all schemes tend toward saturation, suggesting that an appropriate sampling density should be selected to avoid redundant training overhead.

\subsection{The Impact of Obstacle Size}

\begin{figure}[t]
\setlength{\abovecaptionskip}{0pt}
\setlength{\belowcaptionskip}{0pt}
	\centering
    \includegraphics[width=0.95\linewidth]{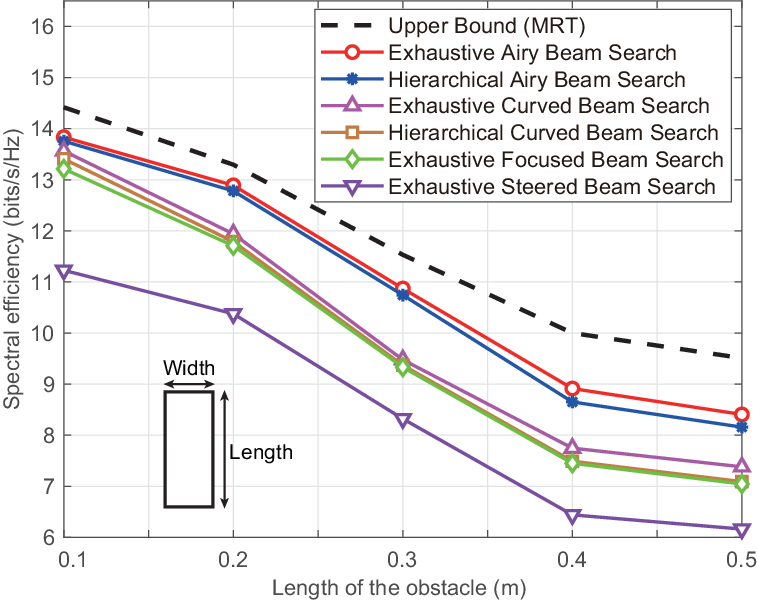}
	\caption{Spectral efficiency vs. length of the obstacle.}
	\label{Fig:height_SP}
\vspace{0em}
\end{figure}

Given a fixed obstacle width of 0.2 m and a length ranging from 0.1 m to 0.5 m, Fig.~\ref{Fig:height_SP} illustrates the spectral efficiency with the obstacle length. As the obstacle dimensions increase, the theoretical upper bound of the achievable rate decreases considerably. This is attributed to the fact that larger obstacles lead to a higher blockage ratio, which subsequently degrades the link quality. Moreover, the spectral efficiency of all beam search schemes declines because the expanding obstacle size enlarges the blind zones where beams cannot effectively reach. While focused and curved beams, as typical near-field waveforms, outperform far-field steered beams, they still exhibit a performance gap compared to Airy beams. These results demonstrate that Airy beams serve as a promising solution to further enhance user rates in near-field obstructed scenarios.

\section{Conclusions}\label{SecVI}

This paper proposes a new waveform termed the near-field Airy beam, which integrates the non-uniform amplitude profile of the Airy function with the non-linear phase response characteristic of the near-field region. The proposed beam's distinctive amplitude and phase responses are specifically tailored to match the inherent characteristics of obstructed near-field channels. By regulating four key parameters of the near-field Airy beam, the propagation trajectory, energy distribution, and focal point can be precisely controlled. This flexibility enables the beam to effectively bypass obstacles while maintaining high energy concentration at the user’s location. To facilitate beam alignment in practical scenarios where the locations of both obstacles and users are unknown, we design an Airy beam codebook along with a low-overhead hierarchical beam training scheme. Moreover, an Airy beamforming algorithm for hybrid beamformer architectures is presented.

Simulation results demonstrate that: 1) In obstructed environments, the proposed Airy beam achieves a received power gain exceeding $3\text{ dB}$ compared to conventional focused and curved beams, while closely approaching the theoretical performance upper bound. 2) In beam training scenarios with unknown obstacles and users, the proposed scheme enhances spectral efficiency over focused and curved beam methods while maintaining a comparable level of training overhead. 3) The proposed Airy beam consistently outperforms existing benchmarks across mmWave to THz bands, and exhibits robust performance across varying obstacle dimensions.

\end{document}